\documentclass[12pt,a4paper]{article}
\makeatletter
\long\def\@makecaption#1#2{%
\vskip\abovecaptionskip%
\sbox\@tempboxa{#1 #2}
\ifdim \wd\@tempboxa >\hsize%
#1 #2\par
\else
\global \@minipagefalse
\hb@xt@\hsize{\hfil\box\@tempboxa\hfil}%
\fi
\vskip\belowcaptionskip}
\makeatother
\begin{document}
\begin{titlepage}
\setcounter{page}{0}
\title{A Lecture on Quantum Logic Gates}
\author{Kazuyuki FUJII\thanks{E-mail address : 
fujii@math.yokohama-cu.ac.jp}\\
Department of Mathematical Sciences\\
Yokohama City University\\
Yokohama, 236-0027\\
Japan}
\date{}
\maketitle
\begin{abstract}
In this note we make a short review of constructions of
n-repeated controlled unitary gates in quantum logic gates.
\end{abstract}
\end{titlepage}
\section{Introduction}
This is one of my lectuers entitled ``Introduction to Quantum 
Computation'' given at Graduate School of Yokohama City University.
The contents of lecture are based on the book [1] and review papers [2], 
[3]. 
The controlled NOT gate (more generally, controlled unitary gates)
plays very important role in quantum logic gates to prove a 
universality.
The constructions of controlled unitary gates or controlled-controlled 
unitary gates are clear and easy to understand.
But the construction of general controlled unitary gates 
(n-repeated controlled unitary gates) seem, in my teaching experience,
not easy to understand for young graduate students.
I thought out some method to make the proof more accessible to them.
I will introduce it in this note.
Maybe it is, more or less, well-known in some field in Pure Mathematics, 
but we are too busy to study such a field leisurely.
I believe that this note will make non-experts more accessible to quantum 
logic gates.
\section{Some Identity on $Z_2$}
Let us start with the mod 2 operation in $Z_2$ : for $x, y \in Z_2$
\begin{equation}
x\oplus y=x+y\ (\makebox{mod} 2).
\end{equation}
From the relations
$$
0\oplus 0=0,\ 0\oplus 1=1,\ 1\oplus 0=1,\ 1\oplus 1=0,
$$
it is easy to see
\begin{equation}
x\oplus y=x+y -2xy,\  \mbox{or}\quad  x+y-x\oplus y=2xy.
\end{equation}
We note that $x\oplus 0=x,\ x\oplus 1=1-x,\ x\oplus x=2x-2x^2=2x(1-x)=0$.

From $x+y-x\oplus y=2xy$ we have
\begin{equation}
x+y+z-(x\oplus y+x\oplus z+y\oplus z)+x\oplus y\oplus z=4xyz
\end{equation}
for $x, y, z \in Z_2$.
The proof is easy, so we leave it to the readers.
Moreover we have
\begin{eqnarray}
&&
x+y+z+w-(x\oplus y+x\oplus z+x\oplus w+y\oplus z+y\oplus w+z\oplus w)
+(x\oplus y\oplus z+\nonumber\\
&&
x\oplus y\oplus w+x\oplus z\oplus w+y\oplus z\oplus w)
-x\oplus y\oplus z\oplus w=8xyzw
\end{eqnarray}
for $x, y, z, w \in Z_2$.
But is this proof so easy ?

Now we define a function
\begin{eqnarray}
F_n(x_1, \cdots, x_n)
&=&
\sum_{i=1}^nx_i-\sum_{i<j}^nx_i\oplus x_j
+\sum_{i<j<k}^nx_i\oplus x_j\oplus x_k - \cdots\nonumber\\
&+&
(-1)^{n-2}\sum_{i=1}^nx_1\oplus\cdots\oplus 
\breve{x_i}\oplus\cdots\oplus x_n
+(-1)^{n-1}x_1\oplus\cdots\oplus x_n
\end{eqnarray}
for $x_1, \cdots, x_n \in Z_2$.
From (2), (3) and (4) we have $F_2(x_1,x_2)=2x_1x_2$ 
and $F_3(x_1,x_2,x_3)=4x_1x_2x_3$ and 
$F_4(x_1,x_2,x_3,x_4)=8x_1x_2x_3x_4$.
From these relations it is easy to conjecture 
{\bf Proposition A}
\begin{equation}
F_n(x_1,x_2,\cdots,x_n)=2^{n-1}x_1x_2\cdots x_n.
\end{equation}
This is well-known [4], but I don't know the usual proof ( in [4] there is 
no proof).
This proof may be not easy for non-experts against the claim in the book 
[1], see pp. 30-31.
Here let us introduce a new (?) method to prove this.
For that we must make some mathematical preparations.
First let us extend the operation $\oplus$ in $Z_2$ to
an operation $\tilde \oplus$ in $Z$ : for $x, y \in Z$
\begin{equation}
x{\tilde\oplus}y\equiv x+y-2xy.
\end{equation}
Of course $x{\tilde\oplus}y=x\oplus y$ for $x, y \in Z_2$.
Here we list some important properties of this operation :
\par \noindent 
{\bf Lemma 1}\quad For $x, y, z \in Z$
\begin{eqnarray}
x\tilde\oplus y &=&
y\tilde\oplus x, \nonumber\\
(x\tilde\oplus y)\tilde\oplus z &=&
x\tilde\oplus(y\tilde\oplus z), \nonumber\\
x\tilde\oplus z+y\tilde\oplus z &=&
(x+y)\tilde\oplus z+z, \nonumber\\
x\tilde\oplus z-y\tilde\oplus z &=&
(x-y)\tilde\oplus z-z
\end{eqnarray}
We also note that $x \tilde\oplus 0=x,\  x \tilde\oplus 1=1 - x,\ 
x \tilde\oplus x=2x(1 - x)$.

What we want to prove in this section is the following recurrent 
relation
\par \noindent  
{\bf Proposition B}\quad For $x_1, \cdots, x_n,x_{n+1} \in Z_2$
\begin{equation}
F_{n+1}(x_1,\cdots,x_n,x_{n+1})=
F_n(x_1,\cdots,x_n)+x_{n+1}-F_n(x_1,\cdots,x_n)\tilde\oplus x_{n+1}.
\end{equation}
If we can prove this , then it is easy to see from the definition of
$\tilde\oplus$
\begin{equation}
F_{n+1}(x_1,\cdots,x_n,x_{n+1})=2x_{n+1}F_n(x_1,\cdots,x_n).
\end{equation}
From this we have Proposition A.
Before giving the proof to Proposition B we make some preliminaries.
\par \noindent 
{\bf Lemma 2}\quad For $x_1, \cdots, x_n,z \in Z_2$
\begin{eqnarray}
&&
\sum_{i=1}^nx_i\oplus z=\left(\sum_{i=1}^nx_i\right)\tilde\oplus 
z+(n-1)z,\\
&&
\sum_{i=1}^n(-1)^{i-1}x_i\oplus z=
\left(\sum_{i=1}^n(-1)^{i-1}x_i\right)\tilde\oplus z
-\frac{1+(-1)^n}2z.
\end{eqnarray}
The proof is straightforward from Lemma 1.
\par \noindent 
{\bf Lemma 3}\quad For $n\geq 2$
\begin{equation}
\sum_{i=1}^{n-1}(-1)^i({}_nC_i-1)=-\frac{1+(-1)^n}2.
\end{equation}
The proof is as follows :
\begin{eqnarray*}
\makebox{Left hand side}
&=&
\sum_{i=1}^{n-1}(-1)^i{_nC_i}+\sum_{i=1}^{n-1}(-1)^{i+1}\\
&=&
\sum_{i=0}^{n}(-1)^i{_nC_i}-\{1+(-1)^n\}+\sum_{i=0}^{n-2}(-1)^i\\
&=&
(1-1)^n-\{1+(-1)^n\}+\frac{1-(-1)^{n-1}}2\\
&=&
-\{1+(-1)^n\}+\frac{1+(-1)^n}2\\
&=&
-\frac{1+(-1)^n}2.\quad \heartsuit
\end{eqnarray*}
First of all let us show my idea to prove Proposition B with a simple 
example.
\begin{eqnarray*}
F_3(x_1,x_2,x_3)
&=&
x_1+x_2+x_3-(x_1\oplus x_2+x_1\oplus x_3 +x_2\oplus x_3)
+x_1\oplus x_2\oplus x_3\\
&=&
x_1+x_2-x_1\oplus x_2+x_3
-\{x_1\oplus x_3+x_2\oplus x_3-x_1\oplus x_2\oplus x_3\}\\
&=&
F_2(x_1,x_2)+x_3
-\left\{(x_1+x_2)\tilde\oplus x_3+x_3-x_1\oplus x_2\oplus x_3\right\}\\
&=&
F_2(x_1,x_2)+x_3
-\left\{(x_1+x_2-x_1\oplus x_2)\tilde\oplus x_3-x_3+x_3\right\}\\
&=&
F_2(x_1,x_2)+x_3-F_2(x_1,x_2)\tilde\oplus x_3.\quad \heartsuit
\end{eqnarray*}
Let us start the proof of Proposition B.
\begin{eqnarray*}
\lefteqn{F_{n+1}(x_1,\cdots,x_n,x_{n+1})
=\sum_{i=1}^{n+1}x_i-\sum_{i<j}^{n+1}x_i\oplus x_j
+\sum_{i<j<k}^{n+1}x_i\oplus x_j\oplus x_k- \cdots}\\
&&
+(-1)^{n-1}\sum_{i=1}^{n+1}x_1\oplus\cdots\oplus\breve{x_i}\oplus\cdots 
\oplus x_{n+1}+(-1)^nx_1\oplus\cdots\oplus x_n\oplus x_{n+1}\\
&=&
F_n(x_1,\cdots,x_n)+x_{n+1}
-\sum_{i=1}^nx_i\oplus x_{n+1}+\sum_{i<j}^nx_i\oplus x_j\oplus 
x_{n+1}- \cdots\\
&&
+(-1)^{n-1}\sum_{i=1}^nx_1\oplus\cdots\oplus\breve{x_i}\oplus\cdots
\oplus x_n\oplus x_{n+1}+(-1)^nx_1\oplus\cdots\oplus x_n\oplus x_{n+1}\\
&=&
F_n(x_1,\cdots,x_n)+x_{n+1}\\
&&
-\left\{\left(\sum_{i=1}^{n}x_i \right)\tilde\oplus x_{n+1}
+({}_nC_1-1)x_{n+1}\right\}\\
&&
+\left\{\left(\sum_{i<j}^{n}x_i\oplus x_j \right)\tilde\oplus x_{n+1}
+({}_nC_2-1)x_{n+1}\right\}\\
&&
\quad \cdots\\
&&
+(-1)^{n-1}\left\{\left(\sum_{i=1}^nx_1\oplus\cdots\oplus\breve{x_i}
\oplus\cdots\oplus x_n\right)\tilde\oplus x_{n+1}
+({}_nC_{n-1}-1)x_{n+1}\right\}\\
&&
+(-1)^nx_1\oplus\cdots\oplus x_n\oplus x_{n+1}\\
&=&
F_n(x_1,\cdots,x_n)+x_{n+1}
-\left(\sum_{i=1}^nx_i\right)\tilde\oplus x_{n+1}
+\left(\sum_{i<j}^nx_i\oplus x_j\right)\tilde\oplus x_{n+1}- \cdots\\
&&
+(-1)^{n-1}\left(\sum_{i=1}^nx_1\oplus\cdots\oplus\breve{x_i}\oplus
\cdots\oplus x_n\right)\tilde\oplus x_{n+1}
+(-1)^nx_1\oplus\cdots\oplus x_n\oplus x_{n+1}\\
&&
+\left\{\sum_{i=1}^{n-1}(-1)^i({}_nC_i-1)\right\}x_{n+1}\\
&=&
F_n(x_1,\cdots,x_n)+x_{n+1}
-\left\{\sum_{i=1}^nx_i-\sum_{i<j}^nx_i\oplus x_j+\cdots\right.\\
&&
\left.+(-1)^{n-2}
\sum_{i=1}^nx_1\oplus\cdots\oplus\breve{x_i}\oplus\cdots\oplus x_n
+(-1)^{n-1}x_1\oplus\cdots\oplus x_n\right\}\tilde\oplus x_{n+1}\\
&&
+\frac{1+(-1)^n}2x_{n+1}-\frac{1+(-1)^n}2x_{n+1}
\qquad
\makebox{by Lemma 2 and Lemma 3}\\
&=&
F_n(x_1,\cdots,x_n)+x_{n+1}-F_n(x_1,\cdots,x_n)\tilde\oplus x_{n+1}.
\quad \heartsuit
\end{eqnarray*}
{\bf One word} : I introduced one method to prove Proposition A.
Of course we have an another one [5], but in my teaching experience my 
method was popular among young graduate students.
\section{General Controlled Unitary Gates}
Let a basis of 1-qubit space $\mbox{\boldmath $C$}^2$ be $\{|0\rangle , 
|1\rangle \}$.,
$$
|0\rangle =
\left(
\begin{array}{c}
1\\
0
\end{array}
\right),
\quad
|1\rangle =
\left(
\begin{array}{c}
0\\
1
\end{array}
\right)
$$
and 2-qubit space $\mbox{\boldmath $C$}^2\otimes \mbox{\boldmath 
$C$}^2$ be
\[
\mbox{\boldmath $C$}^2\otimes \mbox{\boldmath $C$}^2
=\mbox{Vect}_C\{|0,0\rangle,|0,1\rangle,|1,0\rangle,|1,1\rangle\}
\cong \mbox{\boldmath $C$}^4
\]
where $|i,j\rangle\equiv |i\rangle\otimes|j\rangle$ for $i,j=0,1$.

The controlled NOT operation is defined as
\begin{eqnarray}
\mbox{C-NOT} : &\quad&
|0,0\rangle\to |0,0\rangle,\quad |0,1\rangle\to |0,1\rangle, \nonumber\\
&&
|1,0\rangle\to |1,1\rangle,\quad  |1,1\rangle\to |1,0\rangle
\end{eqnarray}
and , therefore, the matrix representation is
\begin{eqnarray}
\mbox{C-NOT}=
\left(
\begin{array}{cccc}
1&0&0&0\\
0&1&0&0\\
0&0&0&1\\
0&0&1&0
\end{array}
\right)
\end{eqnarray}
and represented graphically as
\newpage
\begin{figure}[htb]
\setlength{\unitlength}{1mm}  
\begin{picture}(150,40)
\put(10,35){\line(1,0){50}}   
\put(10,10){\line(1,0){22}}   
\put(38,10){\line(1,0){22}}   
\put(0,30){\makebox(9,10)[r]{$|0\rangle$}} 
\put(0,10){\makebox(9,10)[r]{$|0\rangle$}} 
\put(0,0){\makebox(9,10)[r]{$|1\rangle$}} 
\put(61,30){\makebox(9,10)[l]{$|0\rangle$}} 
\put(61,10){\makebox(9,10)[l]{$|0\rangle$}} 
\put(61,0){\makebox(9,10)[l]{$|1\rangle$}} 
\put(35,13){\line(0,1){22}}     
\put(32,30){\makebox(6,10){$\bullet$}} 
\put(35,10){\circle{6}}               
\put(32,5){\makebox(6,10){X}}         
\put(85,35){\line(1,0){50}}   
\put(85,10){\line(1,0){22}}   
\put(113,10){\line(1,0){22}}   
\put(75,30){\makebox(9,10)[r]{$|1\rangle$}} 
\put(75,10){\makebox(9,10)[r]{$|0\rangle$}} 
\put(75,0){\makebox(9,10)[r]{$|1\rangle$}} 
\put(136,30){\makebox(9,10)[l]{$|1\rangle$}} 
\put(136,10){\makebox(9,10)[l]{$|1\rangle$}} 
\put(136,0){\makebox(9,10)[l]{$|0\rangle$}} 
\put(110,13){\line(0,1){22}}     
\put(107,30){\makebox(6,10){$\bullet$}} 
\put(110,10){\circle{6}}               
\put(107,5){\makebox(6,10){X}}         
\end{picture}
\caption{}
\end{figure}
\noindent
Let $U$ be an arbitrarily unitary matrix in $U(2)$.
Then the controlled unitary gates are defined as
\begin{eqnarray}
\mbox{C-U} : &\quad&
|0,0\rangle\to |0,0\rangle,\quad |0,1\rangle\to |0,1\rangle, \nonumber\\
&&
|1\rangle\otimes|0\rangle\to |1\rangle\otimes(U|0\rangle),\quad  
|1\rangle\otimes|1\rangle\to |1\rangle\otimes(U|1\rangle)
\end{eqnarray}
more briefly,
\begin{equation}
\mbox{C-U} :
|x\rangle\otimes|y\rangle\to |x\rangle\otimes(U^x|y\rangle) \ \
\mbox{for} \ \
x,y \in Z_2
\end{equation}
and represented graphically as
\begin{figure}[htb]
\setlength{\unitlength}{1mm}  
\begin{picture}(150,40)
\put(60,35){\line(1,0){50}}   
\put(60,10){\line(1,0){22}}   
\put(88,10){\line(1,0){22}}   
\put(50,30){\makebox(9,10)[r]{$|x\rangle$}} 
\put(50,5){\makebox(9,10)[r]{$|y\rangle$}} 
\put(111,30){\makebox(9,10)[l]{$|x\rangle$}} 
\put(111,5){\makebox(9,10)[l]{$U^x|y\rangle$}} 
\put(85,13){\line(0,1){22}}     
\put(82,30){\makebox(6,10){$\bullet$}} 
\put(85,10){\circle{6}}               
\put(82,5){\makebox(6,10){U}}         
\end{picture}
\caption{}
\end{figure}

\par \noindent
If $U=\sigma_2=X=
\left(
\begin{array}{cc}
0&1\\
1&0
\end{array}
\right)$,
then the controlled unitary gate is just controlled NOT gate.
The controlled-controlled unitary gates are defined as
\begin{equation}
\mbox{C-C-U} :
|x\rangle\otimes|y\rangle\otimes|z\rangle\to
|x\rangle\otimes|y\rangle\otimes(U^{xy}|z\rangle)\ \
\mbox{for}\ \ x,y,z \in Z_2 .
\end{equation}

The controlled-controlled unitary gates are constructed by making use of 
several controlled unitary gates and controlled NOT gates : 
Let $U$ be an arbitrarily unitary matrix in $U(2)$ and $V$ a unitary one 
in $U(2)$ satisfying $V^2=U$.
Then by relation (2)
\begin{equation}
V^{x+y-x\oplus y}=V^{2xy}=(V^2)^{xy}=U^{xy},
\end{equation}
controlled-controlled $U$ gate is graphically represented as
\begin{figure}[htb]
\setlength{\unitlength}{1mm}  
\begin{picture}(150,60)
\put(10,50){\line(1,0){114}}   
\put(10,30){\line(1,0){54}}   
\put(70,30){\line(1,0){32}}   
\put(108,30){\line(1,0){16}}   
\put(10,10){\line(1,0){16}}   
\put(32,10){\line(1,0){13}}   
\put(51,10){\line(1,0){32}}   
\put(89,10){\line(1,0){35}}   
\put(0,45){\makebox(9,10)[r]{$|x\rangle$}} 
\put(0,25){\makebox(9,10)[r]{$|y\rangle$}} 
\put(0,5){\makebox(9,10)[r]{$|z\rangle$}} 
\put(125,45){\makebox(25,10)[l]{$|x\rangle$}} 
\put(125,25){\makebox(25,10)[l]{$|y\rangle$}} 
\put(125,5){\makebox(25,10)[l]{$U^{xy}|z\rangle$}} 
\put(29,13){\line(0,1){37}}     
\put(48,13){\line(0,1){17}}     
\put(67,33){\line(0,1){17}}     
\put(86,13){\line(0,1){17}}     
\put(105,33){\line(0,1){17}}     
\put(26,45){\makebox(6,10){$\bullet$}} 
\put(45,25){\makebox(6,10){$\bullet$}} 
\put(64,45){\makebox(6,10){$\bullet$}} 
\put(83,25){\makebox(6,10){$\bullet$}} 
\put(102,45){\makebox(6,10){$\bullet$}} 
\put(29,10){\circle{6}}               
\put(48,10){\circle{6}}               
\put(67,30){\circle{6}}               
\put(86,10){\circle{6}}               
\put(105,30){\circle{6}}               
\put(64,25){\makebox(6,10){X}}         
\put(102,25){\makebox(6,10){X}}         
\put(26,5){\makebox(6,10){$V$}}         
\put(45,5){\makebox(6,10){$V$}}         
\put(83,5){\makebox(6,10){$V^{\mbox{\dag}}$}}   
\end{picture}
\caption{}
\end{figure}

The controlled-controlled-controlled unitary gates are constructed by 
the following :
Let $U$ be an arbitrarily unitary matrix in $U(2)$ and $V$ be a unitary 
one in $U(2)$ satisfying $V^4=U$.
Then making use of (3)
\begin{equation}
V^{x+y+z-(x\oplus y+x\oplus z+y\oplus z)+x\oplus y\oplus z}
=V^{4xyx}=U^{xyz},
\end{equation}
controlled-controlled-controlled $U$ gate is graphically represented as
\newpage
\begin{figure}[htb]
\setlength{\unitlength}{1mm} 
\begin{picture}(160,100)
\put(6,80){\line(1,0){144}}   
\put(6,60){\line(1,0){29}}   
\put(41,60){\line(1,0){10}}   
\put(57,60){\line(1,0){50}}   
\put(113,60){\line(1,0){26}}   
\put(145,60){\line(1,0){5}}   
\put(6,40){\line(1,0){53}}   
\put(65,40){\line(1,0){10}}   
\put(81,40){\line(1,0){2}}   
\put(89,40){\line(1,0){10}}   
\put(105,40){\line(1,0){10}}   
\put(121,40){\line(1,0){10}}   
\put(137,40){\line(1,0){13}}   
\put(6,20){\line(1,0){5}}   
\put(17,20){\line(1,0){2}}   
\put(25,20){\line(1,0){2}}   
\put(33,20){\line(1,0){10}}   
\put(49,20){\line(1,0){18}}   
\put(73,20){\line(1,0){18}}   
\put(97,20){\line(1,0){26}}   
\put(129,20){\line(1,0){21}}   
\put(0,75){\makebox(5,10)[r]{$|x\rangle$}} 
\put(0,55){\makebox(5,10)[r]{$|y\rangle$}} 
\put(0,35){\makebox(5,10)[r]{$|z\rangle$}} 
\put(0,15){\makebox(5,10)[r]{$|w\rangle$}} 
\put(151,75){\makebox(9,10)[l]{$|x\rangle$}} 
\put(151,55){\makebox(9,10)[l]{$|y\rangle$}} 
\put(151,35){\makebox(9,10)[l]{$|z\rangle$}} 
\put(151,15){\makebox(9,10)[l]{$U^{xyz}|w\rangle$}} 
\put(14,23){\line(0,1){57}}     
\put(22,23){\line(0,1){37}}     
\put(30,23){\line(0,1){17}}     
\put(38,63){\line(0,1){17}}     
\put(46,23){\line(0,1){37}}     
\put(54,63){\line(0,1){17}}     
\put(62,43){\line(0,1){37}}     
\put(70,23){\line(0,1){17}}     
\put(78,43){\line(0,1){37}}     
\put(86,43){\line(0,1){17}}     
\put(94,23){\line(0,1){17}}     
\put(102,43){\line(0,1){17}}     
\put(110,63){\line(0,1){17}}     
\put(118,43){\line(0,1){17}}     
\put(126,23){\line(0,1){17}}     
\put(134,43){\line(0,1){17}}     
\put(142,63){\line(0,1){17}}     
\put(11,75){\makebox(6,10){$\bullet$}} 
\put(19,55){\makebox(6,10){$\bullet$}} 
\put(27,35){\makebox(6,10){$\bullet$}} 
\put(35,75){\makebox(6,10){$\bullet$}} 
\put(43,55){\makebox(6,10){$\bullet$}} 
\put(51,75){\makebox(6,10){$\bullet$}} 
\put(59,75){\makebox(6,10){$\bullet$}} 
\put(67,35){\makebox(6,10){$\bullet$}} 
\put(75,75){\makebox(6,10){$\bullet$}} 
\put(83,55){\makebox(6,10){$\bullet$}} 
\put(91,35){\makebox(6,10){$\bullet$}} 
\put(99,55){\makebox(6,10){$\bullet$}} 
\put(107,75){\makebox(6,10){$\bullet$}} 
\put(115,55){\makebox(6,10){$\bullet$}} 
\put(123,35){\makebox(6,10){$\bullet$}} 
\put(131,55){\makebox(6,10){$\bullet$}} 
\put(139,75){\makebox(6,10){$\bullet$}} 
\put(14,20){\circle{6}}               
\put(22,20){\circle{6}}               
\put(30,20){\circle{6}}               
\put(38,60){\circle{6}}               
\put(46,20){\circle{6}}               
\put(54,60){\circle{6}}               
\put(62,40){\circle{6}}               
\put(70,20){\circle{6}}               
\put(78,40){\circle{6}}               
\put(86,40){\circle{6}}               
\put(94,20){\circle{6}}               
\put(102,40){\circle{6}}               
\put(110,60){\circle{6}}               
\put(118,40){\circle{6}}               
\put(126,20){\circle{6}}               
\put(134,40){\circle{6}}               
\put(142,60){\circle{6}}               
\put(35,55){\makebox(6,10){X}}         
\put(51,55){\makebox(6,10){X}}         
\put(107,55){\makebox(6,10){X}}         
\put(139,55){\makebox(6,10){X}}         
\put(59,35){\makebox(6,10){X}}         
\put(75,35){\makebox(6,10){X}}         
\put(83,35){\makebox(6,10){X}}         
\put(99,35){\makebox(6,10){X}}         
\put(115,35){\makebox(6,10){X}}         
\put(131,35){\makebox(6,10){X}}         
\put(11,15){\makebox(6,10){$V$}}         
\put(19,15){\makebox(6,10){$V$}}         
\put(27,15){\makebox(6,10){$V$}}         
\put(43,15){\makebox(6,10){$V^{\mbox{\dag}}$}}         
\put(67,15){\makebox(6,10){$V^{\mbox{\dag}}$}}         
\put(91,15){\makebox(6,10){$V^{\mbox{\dag}}$}}         
\put(123,15){\makebox(6,10){$V$}}        
\end{picture}
\caption{}
\end{figure}
\noindent
The general controlled unitary gates are constructed by the following :
Let $U$ be an arbitrarily unitary matrix in $U(2)$ and $V$ be a unitary 
one in $U(2)$ satisfying $V^{2^{n-1}}=U$. Then making use of relation (6)
\begin{equation}
V^{F_n(x_1,x_2,\cdots,x_n)}
=V^{2^{n-1}x_1 x_2 \cdots x_n}=U^{x_1 x_2 \cdots x_n},
\end{equation}
the construction of n-repeated controlled $U$ gate is as follows : 
For example the block implementing $V^{x_i\oplus x_j\oplus x_k}$ is 
graphically constructed as
\newpage
\begin{figure}[htb]
\setlength{\unitlength}{1mm}
\begin{picture}(150,105)
\put(0,94){\makebox(20,10)[r]{$|x_1\rangle$}}    
\put(25,99){\line(1,0){100}}                     
\put(130,94){\makebox(20,10)[l]{$|x_1\rangle$}}  
\put(0,74){\makebox(20,10)[r]{$|x_i\rangle$}}     
\put(25,79){\line(1,0){100}}                     
\put(40,74){\makebox(5,10){$\bullet$}}            
\put(42.5,61.5){\line(0,1){18}}                   
\put(100,74){\makebox(5,10){$\bullet$}}           
\put(102.5,61.5){\line(0,1){18}}                  
\put(130,74){\makebox(20,10)[l]{$|x_i\rangle$}}   
\put(0,54){\makebox(20,10)[r]{$|x_j\rangle$}}     
\put(42.5,59){\circle{5}}
\put(40,54){\makebox(5,10){X}}                    
\put(55,54){\makebox(5,10){$\bullet$}}            
\put(57.5,41.5){\line(0,1){18}}                   
\put(85,54){\makebox(5,10){$\bullet$}}            
\put(87.5,41.5){\line(0,1){18}}
\put(102.5,59){\circle{5}}                        
\put(100,54){\makebox(5,10){X}}                   
\put(25,59){\line(1,0){15}}                       
\put(45,59){\line(1,0){55}}                       
\put(105,59){\line(1,0){20}}                      
\put(130,54){\makebox(20,10)[l]{$|x_j\rangle$}}   
%
\put(0,34){\makebox(20,10)[r]{$|x_k\rangle$}}     
\put(57.5,39){\circle{5}}  
\put(55,34){\makebox(5,10){X}}                    
\put(70,34){\makebox(5,10){$\bullet$}}            
\put(72.5,7.5){\line(0,1){32}}                    
\put(87.5,39){\circle{5}}
\put(85,34){\makebox(5,10){X}}                           
\put(25,39){\line(1,0){30}}                       
\put(60,39){\line(1,0){25}}                       
\put(90,39){\line(1,0){35}}                       
\put(130,34){\makebox(20,10)[l]{$|x_k\rangle$}}   
%
\put(0,14){\makebox(20,10)[r]{$|x_n\rangle$}}     
\put(25,19){\line(1,0){100}}                      
\put(130,14){\makebox(20,10)[l]{$|x_n\rangle$}}   
%
\put(0,0){\makebox(20,10)[r]{$|0\rangle$}}        
\put(72.5,5){\circle{5}}                          
\put(70,0){\makebox(5,10){V}}                     
\put(25,5){\line(1,0){45}}                        
\put(75,5){\line(1,0){50}}                        
\put(130,0){\makebox(25,10)[l]{$V^{x_i\oplus x_j\oplus x_k}|0\rangle$}} 
\put(25,94){$\cdot$}\put(25,89){$\cdot$}\put(25,84){$\cdot$}
\put(25,74){$\cdot$}\put(25,69){$\cdot$}\put(25,64){$\cdot$}
\put(25,54){$\cdot$}\put(25,49){$\cdot$}\put(25,44){$\cdot$}
\put(25,34){$\cdot$}\put(25,29){$\cdot$}\put(25,24){$\cdot$}
\end{picture}
\caption{}
\end{figure}
By combining these blocks like Figure 3 and Figure 4 
we have the n-repeated controlled unitary gates.
But as emphasized in [4] this construction is not efficient.

\vspace{1cm}
\noindent
{\it Acknowledgment.}
The author wishes to thank Michiko Kasai for tex-typing of several 
figures and Tatsuo Suzuki for helpful suggestions.
\end{document}